\documentstyle[12pt,amssym,aasms4]{article}

\lefthead{Fernandes, Ara\'ujo, Pereira \& Landaberry}

\righthead{A Spectral Atlas of the Pre-WN Candidate HD 326823}

\begin{document}

\title{A Spectral Atlas of the Pre-WN Candidate HD 326823}
\altaffiltext{1} {Based on observations mainly done with the 1.52m 
telescope at 
the European Southern Observatory (La Silla, Chile), under the agreement 
with 
the 
Observat\'orio Nacional-MCT (Brazil) and also with the 1.6m telescope at 
the 
Laborat\'orio Nacional de Astrof\' \i sica (Bras\'opolis, Brazil).}

\vskip 1truecm
\author{Marcelo Borges Fernandes}
\affil{borges@dage0.on.br \\} 
\author{Francisco Xavier de Ara\'ujo}
\affil {araujo@on.br \\} 
\author{Claudio Bastos Pereira}
\affil{claudio@on.br \\}
\author{Sayd J. Codina Landaberry}
\affil{sayd@on.br \\
Observat\'orio Nacional, Rua General Jos\'e Cristino 77 \\
CEP 20921-400, Rio de Janeiro  BRAZIL \\}

\begin{abstract}

HD 326823 is a peculiar emission line object. In the literature it was 
considered 
as a pre-WN candidate. Based mainly on high resolution data, 
obtained 
with the FEROS spectrograph, we made an atlas covering 3800-9200 \AA \ region. 
This atlas confirms the presence of several strong He\,{\sc i} 
lines and numerous N lines. There is also a deficiency of H, indicating 
that 
this object is indeed in an evolved stage. In addition, with low resolution 
spectra, we obtained intensities of several lines.

\end{abstract}


\eject
\section{Introduction}

The evolution of massive stars (M$_{ZAMS}$ $>$ 10 M$_\odot$) is not yet 
completely understood. We know that they start their lives as 
O or B 
types and finish them in supernova explosions. Phases like LBV, WN, WC and 
B[e]sg need a 
better comprehension. The existence of few objects in these classes is a 
problem. This problem becomes bigger for transition evolutionary phases  
due to even smaller number of members. The possibility that HD 326823 
(Hen 3-1330) 
is an 
object in a transition phase is the main reason that motivated the present 
study. 

HD 326823 (l$^{II}$ = 344.3;b$^{II}$ = -1.2) belongs to an association of 
distant 
OB supergiants in the direction of the galactic center at a distance R 
$\sim$ 2 
kpc (McGregor, Hyland, \& Hillier 1988). Feast et al. (1961) labelled 
it as a
peculiar 
emission object, with emission lines of H, He\,{\sc i}, Fe\,{\sc ii}, 
Mg\,{\sc ii} and interstellar absorption lines of
Ca\,{\sc ii}. Stephenson (1974) noted that the great strength of the 
He\,{\sc i} emission 
could be attributed to He overabundance and H deficiency.

McGregor et al. (1988) found that the continuum was essentially 
flat 
in flux between 1 and 2.5 $\mu$m, with several He\,{\sc i} emission lines 
and a 
few weak 
H\,{\sc i} emission lines present. They concluded that He overabundance in 
the 
envelope of this star would suggest that HD 326823 is becoming 
a WR 
star. In the UV, Shore et al. (1990) reported a B-type absorption 
spectrum 
consistent with a B1.5 spectral type. In addition, they found that the 
spectrum is 
similar to the one of LBV S128/LMC = R127. 

Lopes, Damineli, \& de Freitas Pacheco (1992), based on Coud\'e data 
limited to few spectral regions, confirmed the result of Carlson \& Henize 
(1979), finding complex 
He\,{\sc i} 
profiles with a double structure. 
This 
structure indicates that the lines are probably formed in different 
circumstellar regions 
with 
rotation. The red Ca\,{\sc ii} 
triplet and the Fe\,{\sc ii} 9997 \AA \ feature also showed 
similar 
two-peak profiles. 
The spectra of Lopes et al. (1992) also showed rather strong N lines, 
indicating 
a 
possible 
overabundance of 
this element. They suggested that HD 326823 may be the 
precursor 
of a WN star.

Sterken et al. (1995) made a spectroscopic and photometric study and 
confirmed 
the very peculiar nature of the star. They also suggested that HD 326823 
may be 
a LBV in a quiescent state, having had S Dor-type eruptions in the past, 
and 
that it is now possibly moving towards a WN configuration. On the other hand, 
Lamers et 
al. (1998) 
and de 
Winter \& P\'erez (1998) included it in the group of unclassified B[e] 
stars.

More recently Walborn \& Fitzpatrick (2000) presented a digital atlas of 20
high-luminosity, peculiar OB spectra in the 3800-4900 \AA \ range. They put HD
326823 in the group of ``Iron Stars'', but they cited that it ``has perhaps the
most bizarre spectrum in this atlas''. Moreover they have emphasized ``the
extreme weakness of all its spectral features, including hydrogen, which must
be very deficient''.

In order to contribute to a better comprehension
of the evolutionary stage of  HD 
326823 , we present in this paper a spectral atlas in the 
3800-9200 \AA \ wavelength range. This atlas is based mainly on high 
resolution data obtained with the FEROS spectrograph (Kaufer et al. 1999) 
attached to the 1.52m ESO telescope.

\section{Observations \& Reductions}

High and low resolution observations were both obtained with the ESO 1.52m 
telescope in La Silla (Chile). 

HD 326823 was observed in high resolution mode on 1999 
June 24, with an exposure time of 1800 seconds, in order 
to present a detailed line list. These data were also used to deblend some 
features present in the low resolution spectra. The Fiber-fed Extended Range 
Optical Spectrograph (FEROS) is a bench-mounted Echelle spectrograph with the
fibers located at the Cassegrain focus with a spectral resolving power
of R = 48\,000 corresponding to 2.2 pixels of 15$\mu$m and with a
wavelength coverage from 3600\AA\, to 9200\AA. FEROS has a complete automatic 
on-line 
reduction. Equivalent widths have been measured using an IRAF task that 
computes the line area above the adopted continuum. The FEROS spectrum of HD 
326823 of the present paper has, in the 5500 \AA \ region, a S/N ratio in the 
continuum of 
aproximately 45.

Low resolution spectra of HD 326823 were taken using a Boller \& 
Chivens 
spectrograph at the 
Cassegrain focus on 1998 March 14, 1998 July 12, and 2000 June 9. Spectra 
of five other similar objects, that are mentioned below, were also 
obtained on different dates. 
Table 1 shows the log of the observations. Two instrumental setups 
were 
employed. The first one made use of grating \#23 with 600 l mm$^{-1}$, 
providing a 
resolution of $\sim$ 4.6 \AA \ in the ranges 3600-7400 \AA \ and 
3800-8700 \AA. The 
other one used grating \#32 with 2400 l mm$^{-1}$, 
providing a 
resolution of $\sim$ 1.0 \AA, in the range 3060-4060 \AA. The 
Cassegrain spectra (hereafter the data obtained with the Boller \& Chivens
spectrograph will be denoted as Cassegrain data) were reduced using standard
IRAF tasks, such as bias  subtraction,  flat-field normalization, and
wavelength calibration. We have done flux  calibrations, and extinction
corrections with the E(B-V) values taken from the  literature. 
Table 2 shows the object designations (Column 1), the E(B-V) values 
(Column 2), and their respective references (Column 3). Spectrophotometric 
standards 
from 
Hamuy et al. (1994) were observed, too.

In the linearized spectra the line intensities have 
been 
measured by the conventional method, adjusting a gaussian function to the 
profile. Uncertainties in the line intensities come mainly from the 
position 
of the 
underlying continuum. We estimate the errors in the measures to be about 
20\% for the
weakest lines (line fluxes $\approx$ 10 on the scale of H$\beta$ = 100) 
and 
about 10\% for the strongest lines. 

In the 5500 \AA \ 
region, the S/N 
ratio in the continuum is aproximately 55 for the Cassegrain 
spectrum 
obtained on 1998 March 14 and 50 for the one obtained 
on 
2000 June 9. In the 3600 \AA \ region, the S/N ratio for the Cassegrain 
spectrum 
is aproximately 35. Since there is no completely line free 
region in 
the spectrum, the S/N derived is an upper limit.

In addition to the FEROS and Cassegrain data, we also obtained spectra 
with medium resolution, using 
a Coud\'e spectrograph. These Coud\'e data came from the 1.6m telescope at the
Laborat\'orio Nacional de Astrof\' \i sica, Bras\'opolis 
(Brazil). We obtained spectra for HD 326823 and six other objects using 
an instrumental setup with a grating with 600 l mm$^{-1}$, providing a 
resolution of $\sim$ 0.8 \AA \ in the range 9800-10200 \AA. The 
data 
reduction has been performed 
using standard IRAF tasks, too. Table 3 shows the log of the observations.

\section{The Spectral Atlas}

In order to identify the lines, we have used the line lists provided by 
Moore 
(1945), 
Thackeray (1967), and Landaberry, Pereira, \& Ara\'ujo (2001). We also 
looked up 
two sites on the web: NIST Atomic Spectra Database Lines Form 
(URL physics.nist.gov/cgi-bin/AtData/lines\_form) and The Atomic 
Line List v2.04 (URL www.pa.uky.edu/~peter/atomic/). 

Tables 4 and 5 show the 
observed wavelength (Column 1), the proposed identification (Column 2), and 
the measured line equivalent width (W) (Column 3) for emission and 
absorption 
lines, respectively, present in the HD 326823 FEROS spectrum. 
There are 10 lines that remain unidentified and they are labelled as ''Uid`` 
in 
Column 2. Figures 1a, 1b, and 1c present 
this FEROS spectrum; the strongest lines (W $>$ 1 \AA) are marked.

Table 6 presents the observed wavelength (Column 1), the proposed 
transitions (Column 2), and 
intensities in ergs cm$^{-2}$ s$^{-1}$ (I($\lambda$)) (Column 3) for emission 
lines 
in the low resolution spectrum in the 
spectral range 3060-4060 \AA. We note that due to the low 
efficiency of the detector in the violet region, we were able to identify only 
a few lines between 3600 and 4060 \AA.

Tables 7 and 8 concern the emission and absorption lines respectively, 
present 
in the low resolution spectra with spectral 
ranges 3600-7400 \AA \ and 3800-8700 \AA. Columns 1 
and 2 in 
the tables show the observed wavelengths and the measured line intensities, 
respectively, for the spectrum taken on 1998 March 14. Columns 3 and 4 
show the 
same data for the spectrum taken on 2000 June 9. Finally, Column 5 
presents the 
proposed identifications. The intensities shown in these two tables are 
relative to 
H$\beta$ = 100 and as H$\beta$ is in emission, we decided to put the minus 
signal before the intensity values of absorption lines. 
We 
found that many emission lines are more intense than H$\beta$, especially 
He\,{\sc i} lines. 
In general, there is a rather significant difference between 
line 
intensities 
of the two dates. In most cases there was an increase between the 
spectrum taken on 1998 March 14 and that on 2000 June 9. On the contrary, 
no 
significant change is seen in the continuum (Figure 2). With the help of the
FEROS spectrum, we have been able to deblend, for instance, the feature 
observed at 5046.9 \AA, estimating the contributions of the
N\,{\sc ii} 
5045.1 \AA, Si\,{\sc ii} 5047.3 \AA, and 
He\,{\sc i} 5047.7 \AA \ lines. These values are shown in Table 7.

The column labelled ``Identification'' in the tables has the element 
transition, the multiplet, and the rest wavelength of the transition. It 
is 
important to say that since differents sources were consulted, it is 
possible 
that more than one ion can be allocated to a single feature. In these 
cases, we 
give some possible alternative identifications. 

Our spectra confirmed, in the optical region, the predominance of He\,{\sc i}, 
Fe\,{\sc ii}, and N\,{\sc ii} lines, and the weakness of H transitions. In 
the next 
section we will comment on the most interesting features and measurements 
obtained 
with FEROS, Cassegrain, and Coud\'e data and discuss the nature of HD 326823.

\section {Discussion}

\subsection {FEROS Spectrum}

Thanks to its high resolution, the FEROS spectrum leads to the identification 
of a great number of 
lines and to a better definition of their profiles. It also allowed us to 
obtain individual equivalent widths of features blended in the Cassegrain 
spectra and so to evaluate the intensity of 
each line.

He\,{\sc i} 3889 \AA, 4922 \AA, 5016 \AA, 5876 \AA, 6678 \AA, 
7065 \AA 
\ and 7281 \AA \ transitions are among those with the highest equivalent 
widths. 
These lines 
present a 
two-peak structure  with blue and red edges reaching 200 km s$^{-1}$ 
(Figure 3). The He\,{\sc i} 3889 \AA \ central absorption is very strong; it
suggests the effect of diluted radiation (see e.g. Niemela \& Sahade 1980).

Regarding H lines, we are only sure of the identification of H$\alpha$ 
and H$\beta$. The presence of P17 is uncertain. H$\alpha$ and H$\beta$ 
also reveal a two-peak structure (Figure 4). It is important to compare 
the 
strength of He and H lines. For instance, He\,{\sc i} 4713 \AA \ 
is 
stronger than H$\beta$ while He\,{\sc i} 6678 \AA \ is 
comparable to 
H$\alpha$. This behavior is confirmed by the low resolution spectra (see 
Table 7).

N lines are numerous and conspicuous. We have identified several permitted 
N\,{\sc i} emission lines as well as permitted and forbidden N\,{\sc ii} 
transitions. It 
is curious that the 
majority of the 
permitted N\,{\sc ii} lines are in absorption. This seems to indicate a 
photospheric origin for them. The forbidden lines are all in emission, 
indicating 
that they 
are formed in the star's envelope. In Figure 5 we can see some N\,{\sc 
ii} profiles.

We have found many permitted and forbidden emission lines attributed to 
low 
ionization 
metals, specially Fe\,{\sc ii}. However, these lines are not very strong. 
Others 
elements 
identified 
are Si\,{\sc ii}, O\,{\sc ii}, and Cr\,{\sc ii}. We also found that two lines 
of 
the red Ca\,{\sc ii} triplet have a high equivalent width and show the two-peak 
structure, confirming 
Lopes 
et al. (1992). Unfortunately, we cannot measure the third line's (8540 
\AA) equivalent width due to the existence of a gap in this spectral region. In 
addition, we 
identified some interstellar absorption bands (4430 
\AA, 
5780 
\AA, 5797 \AA) and also IS lines due to Ca\,{\sc ii} (3934 \AA \ 
and 3968 \AA), 
CH$^+$ 
(3958 \AA, 4233 \AA) and Na I (5890 
\AA, 5896 \AA).

\subsection {Cassegrain Spectra}

The importance of the low resolution spectra is to show the continuum 
behavior 
and to allow the measurement of line intensities, thanks to the flux 
calibration. 
Below, the spectra in two 
different 
wavelength ranges are briefly described. A comparison with other stars 
observed 
by us is presented, too.   

\subsubsection{3060 \AA \ - 4060 \AA}

This spectrum has a wavelength range of $\sim$ 1000 \AA \ and it shows the 
behavior of 
the continuum from the UV to the blue region. We were able to identify only a 
few lines 
in this region due to the noise. The most remarkable feature is the He\,{\sc i} 
3888.7 \AA. In Figure 6, we show the spectra of HD 326823, HD 316285, and HD 
327083. The 
HD 326823 
spectrum in this figure is the one obtained on 1998 July 12. The HD 316285
spectrum shows many lines of the Balmer series, with  P-Cygni profiles. In 
HD 327823 these lines are in 
absorption. 
So, once more we note clearly a H deficiency in HD 326823. 

\subsubsection{3600 \AA \ - 7400 $\AA$ and 3800 \AA \ 
- 8700 \AA}

These spectra cover a wavelength range of $\sim$ 4000 \AA \ and they show 
the 
behavior 
of the continuum from the blue to the red region. 

Figure 7 presents a comparison between HD 326823 and five other 
stars: 
HD 87643, classified as B[e]sg (Miroshnichenko 1998; Oudmaijer 1998 \& 
Oudmaijer et al. 1998) and as Iron Star (Walborn \& Fitzpatrick 2000); HD
316285, classified as LBV (Hillier et al. 1998) and as Iron Star 
(Walborn \& Fitzpatrick 2000); 
HD 327083, classified as LBV or B[e]sg (Machado, de Ara\'ujo, \& 
Lorenz-Martins 
2001); $\eta$ Car, classified as LBV (Walborn 1995); and HR Car, classified 
as 
LBV, too (Hutsemekers \& van Drom 1991). The 
HD 326823 
spectrum in this figure is the one obtained on 1998 March 14. H$\alpha$, 
H$\beta$, and the stronger He\,{\sc i} lines are marked. We can see 
in this figure that only HD 326823 has He\,{\sc i} lines more 
intense than 
H$\alpha$ and 
H$\beta$. These data contribute to the assertion 
that 
HD 326823 is in an advanced evolutionary stage.

\subsection {Coud\'e Spectrum}

Figure 8 shows our Coud\'e spectra of HD 326823 and five other similar 
objects. 
HD 326823 is the only one
that shows the Fe\,{\sc ii} 9997 \AA \ transition (with two-peak structure) 
much 
more intense than P$\delta$. Moreover, the He\,{\sc i} 
10031 \AA \ line is at least as strong as P$\delta$, confirming the H 
deficiency 
in HD 326823.

\subsection{The nature of HD 326823}

Our data confirm that HD 326823 exhibts remarkably strong He\,{\sc i}
lines. On the contrary, the few H\,{\sc i} transitions seen are unusually
weak. The He/H line intensity ratios provide clear evidence for He
overabundance. Such a conclusion is reinforced by the comparison made in
previous subsections with some objects generally classified as B[e]
supergiants (B[e]sg), Luminous Blue Variables (LBV) or ``Iron Stars'' (Lamers
et al. 1998; Walborn \& Fitzpatrick 2000). Moreover our spectra reveal a
great number of conspicous N\,{\sc ii} transitions. N\,{\sc i} emissions and
(perhaps) few N\,{\sc iii} absorption lines may also be seen, but the dominant
ion of this species is surely N\,{\sc ii}. On the other hand, C lines are not seen at
all. Concerning oxygen, only the O\,{\sc i} 4661.2 \AA \ and the O\,{\sc ii}
6046 \AA \ transitions seem to be present in emission. The identification of
O\,{\sc ii} 7319 \AA \ to the feature seen at 7321.8 \AA \ is rather doubtful, since
the contribution of the Fe\,{\sc ii} 7320.7 \AA \ (m7) is probably much more
important (we note that other Fe\,{\sc ii} lines of the same multiplet are
present). Thus, it is likely that N is overabundant too. These conclusions
point in favour of considering HD 326823 as an evolved object.

Let us compare now HD 326823 with other peculiar, evolved massive stars. Due
to the absence of the N\,{\sc iv} 4058 \AA, the N\,{\sc iii} 4630-4634 \AA, and the 
He\,{\sc ii} 4686 \AA \ emissions, it can not be considered as a WNL or WN9ha
star (Walborn \& Fitzpatrick 2000). For the same reason it may not be
classified as Ofpe/WN9 (Walborn 1982). As N lines are concerned, the spectra
of HD 326823 is more or less similar to those of some WNVL stars (Walborn \&
Fitzpatrick 2000). However, in such objects the H\,{\sc i} recombination lines
are the most prominents. Finally, the group of ``Iron Stars'', sometimes cited
in the literature, is not very useful since the objects generally put in this
category have diverse physical natures. We are tempted to conclude that HD
326823 is an unique peculiar star. 

In the last years several works have been devoted to massive stars in
transition. From optical spectroscopy of Ofpe/WN9 and related stars, Nota et
al. (1996) suggested that Of, Ofpe/WN9, LBV, and B[e] stars must be
considered within the same evolutionary scenario. On the basis of infrared
spectra, Morris et al. (1996) concluded that Of supergiants, Ofpe/WN9, WNL,
B[e] supergiants, and LBV, are transitional in their morphological
classification and may be related in their evolution. Some years before
Schulte-Ladbeck et al. (1993) proposed that the distinction between the B[e]
supergiants and other transition objects is a stronger degree of nonspherical
symmetry in their envelopes, probably due to the higher rotational velocities.
We note the apparent disk line profiles seen in our data, which suggest that
HD 326823 may have been a rapid rotator.

From the theoretical point of view two different scenarios have been proposed
to describe the stages between O and WR stars. Crowther et al. (1995)
suggested that the WNL phase follows directly from the Of stars, at least for the 
objects with the highest initial masses. The WNL would then evolve to WNE and
finally WC stars. Langer et al. (1994) proposed the following sequence:
O $\rightarrow$ Of $\rightarrow$ H-rich WN $\rightarrow$ LBV $\rightarrow$ H-poor
WN $\rightarrow$ H-free WN $\rightarrow$ WC $\rightarrow$ SN. In this picture, HD
326823 could be in a post-LBV stage , close to the H-poor WN phase. Such a
statement is hampered by the absence of an abundance analysis, that is beyond
the scope of the present paper.

In summary it is certain that HD 326823 is an indeed peculiar object, which is
in an evolved evolutionary phase of the massive stars. It might be the only
known example of a very precise, short-lived stage. However, its real nature
can not be ascribed until we have accurate determinations of some physical
parameters as the mass-loss rate and the helium and nitrogen abundances. 

The authors thank the useful comments and suggestions of an anonymous
referee, which improved the present paper. M.B.F. acknowledges financial
support from CAPES-Brazil (PhD studentship).

\clearpage

\figcaption[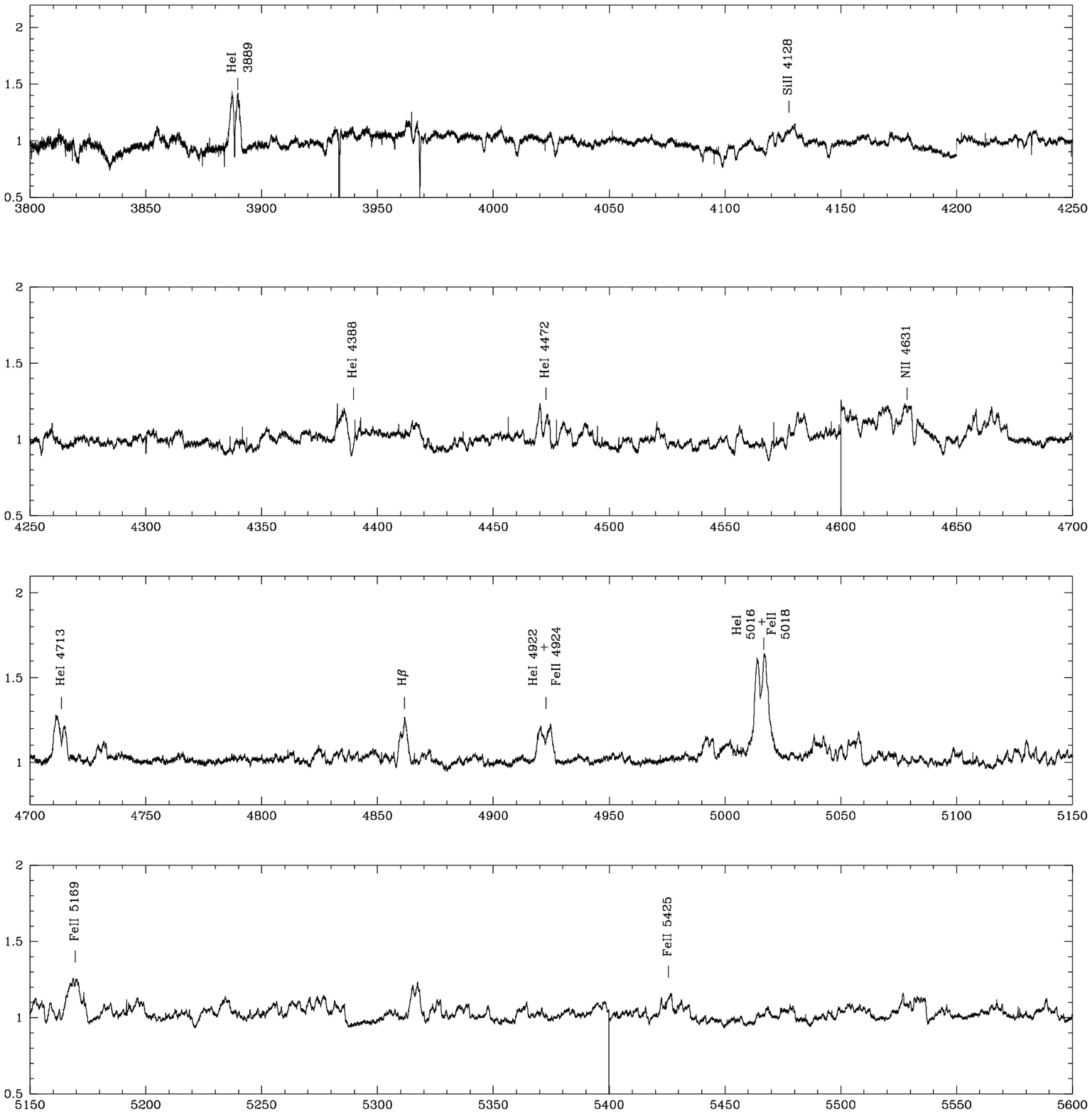]{The spectral atlas from 3800 to 5600 \AA \ 
obtained from the FEROS spectrum. The strongest lines (W $>$ 1 \AA) are marked.
\label{fig1a}}

\figcaption[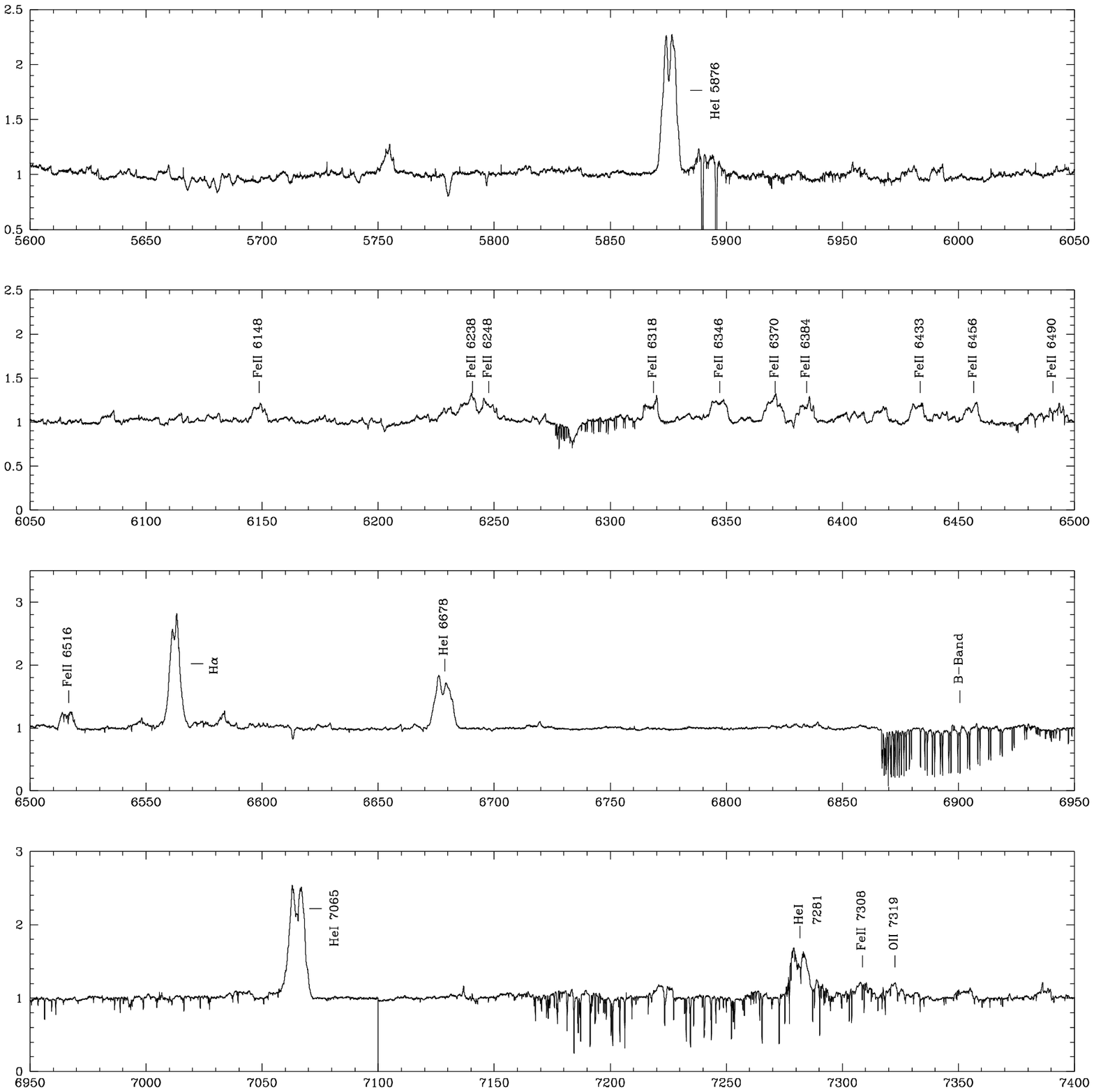]{The spectral atlas from 5600 to 7400 \AA \ 
obtained from the FEROS spectrum. The strongest lines (W $>$ 1 \AA) are marked.
\label{fig1b}}

\figcaption[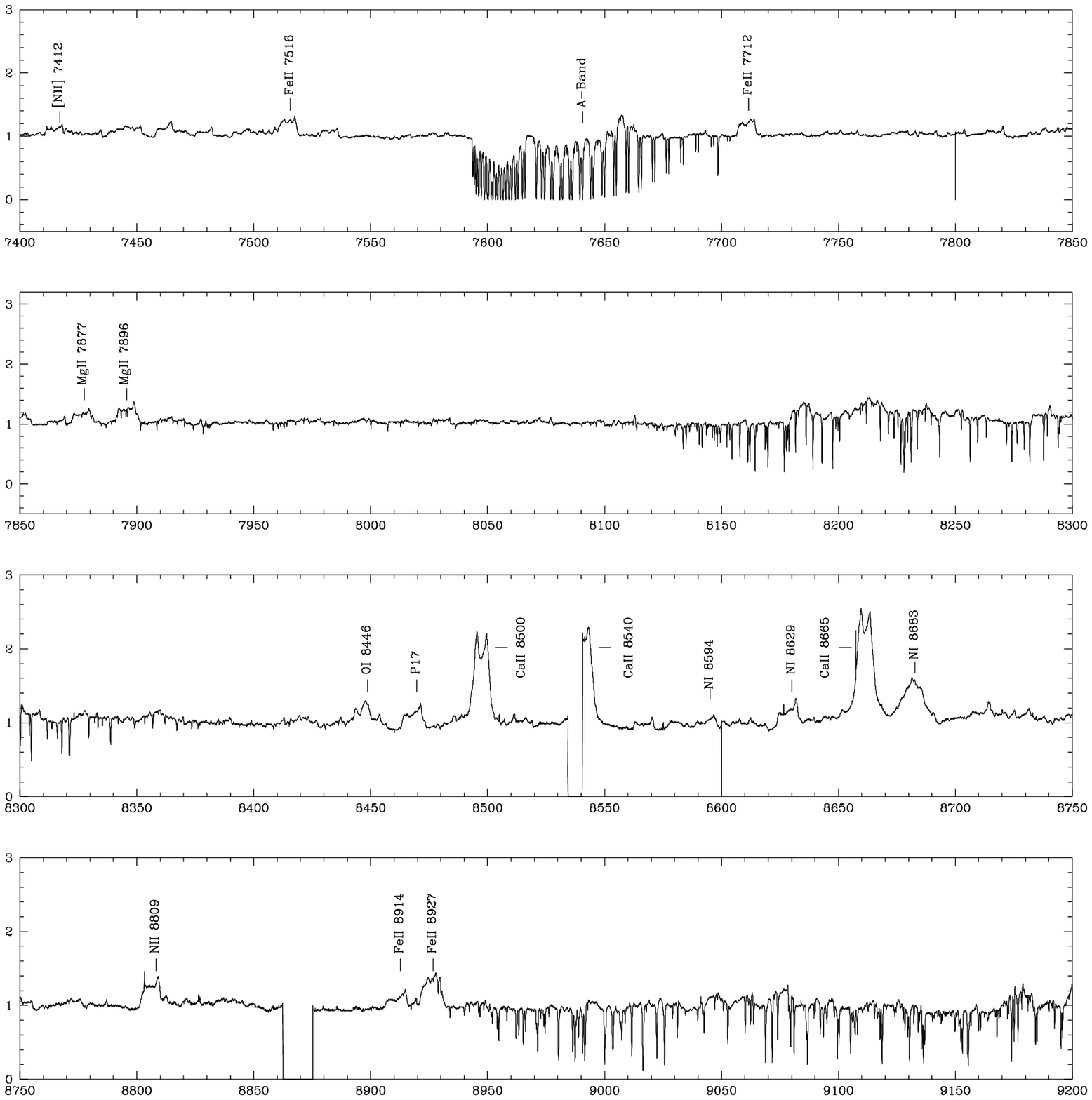]{The spectral atlas from 7400 to 9200 \AA \ 
obtained from the FEROS spectrum. The strongest lines (W $>$ 1 \AA) are marked.
\label{fig1c}}
 
\figcaption[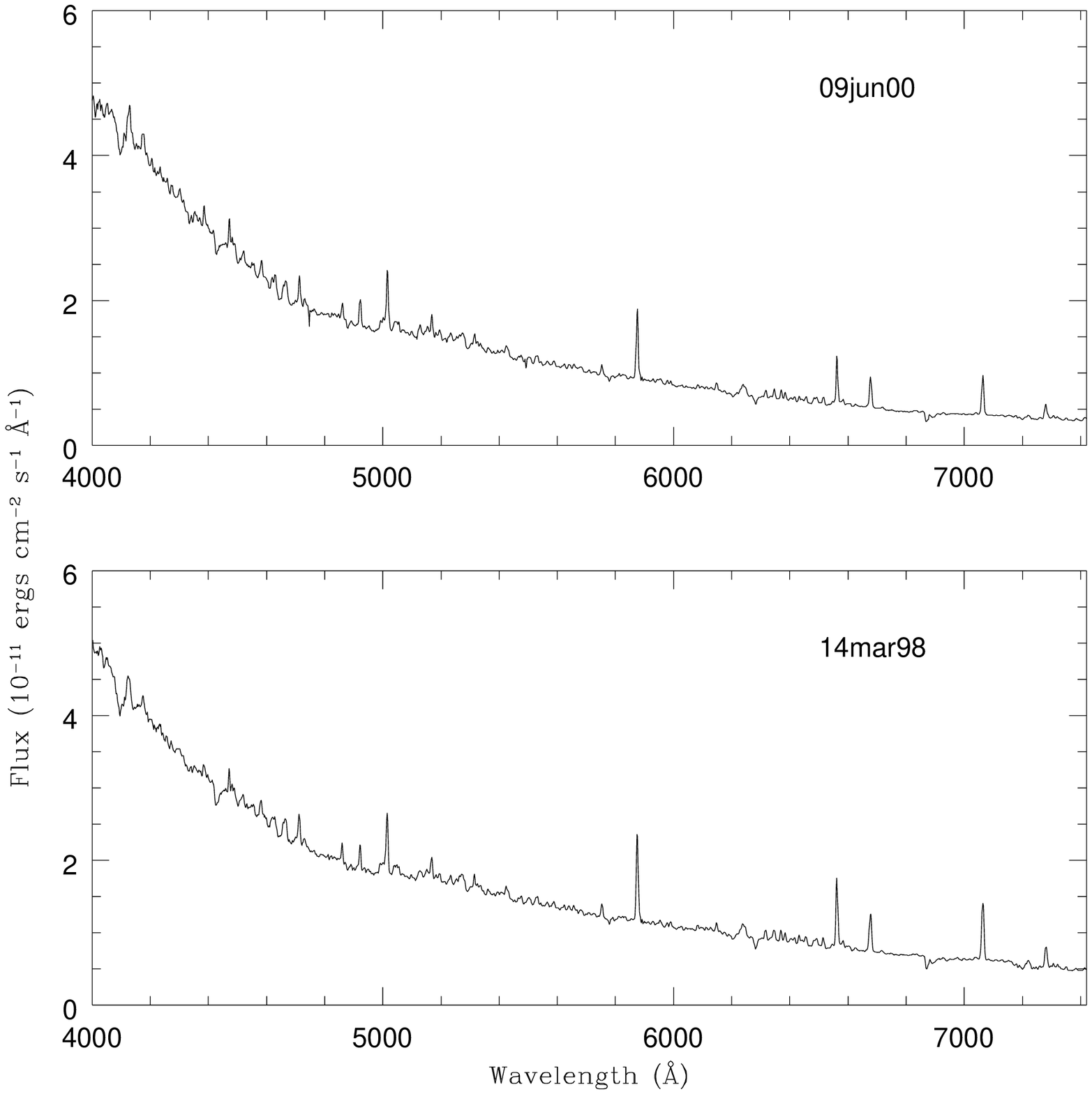]{The HD 326823 low resolution spectra, obtained on 
1998 March 14 and on 2000 June 9. These spectra were corrected for 
extinction, using E(B-V) values.
\label{fig2}} 

\figcaption[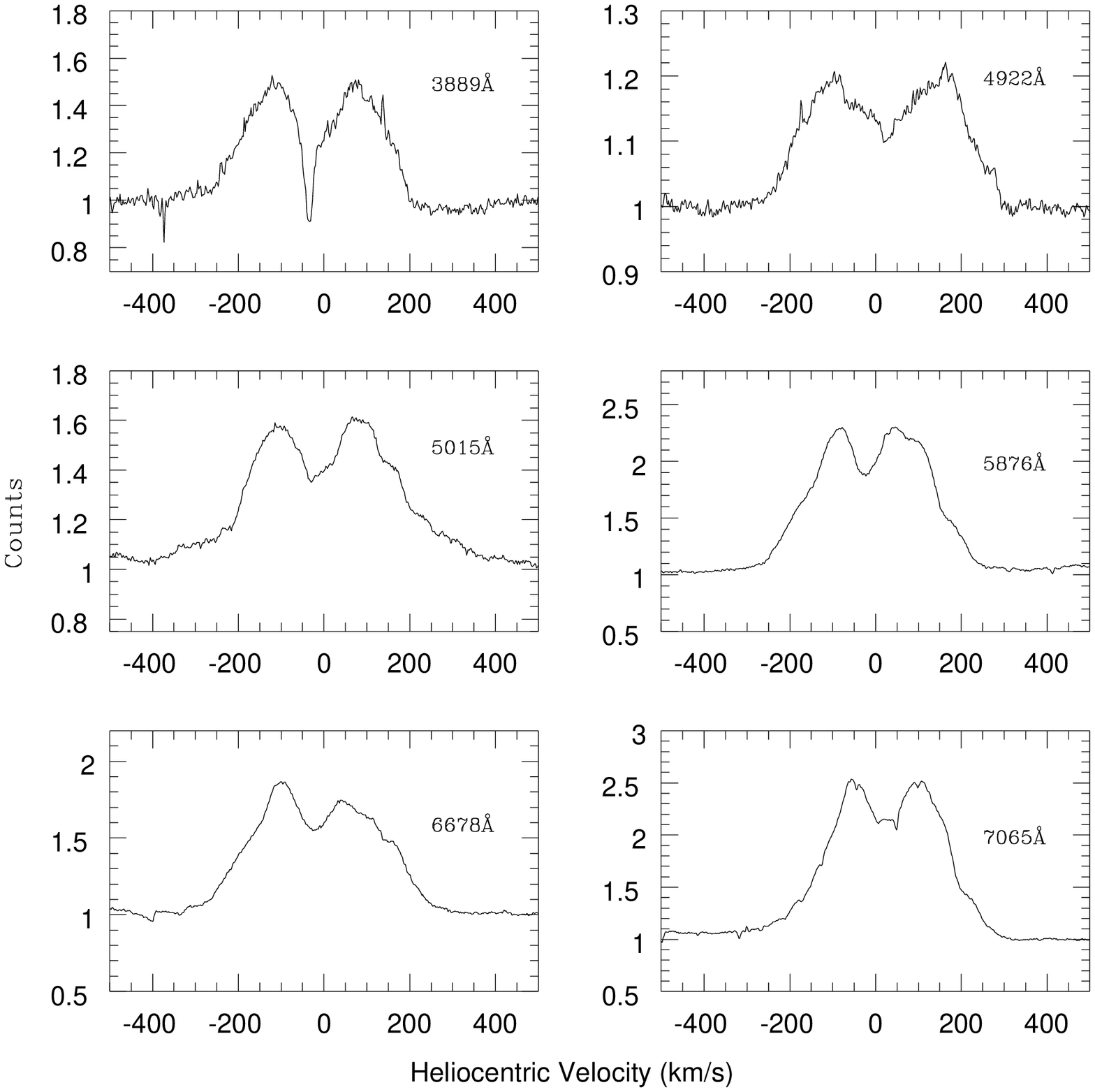]{Profiles of six He\,{\sc i} lines presented in 
the FEROS spectrum.   
\label{fig3}} 

\figcaption[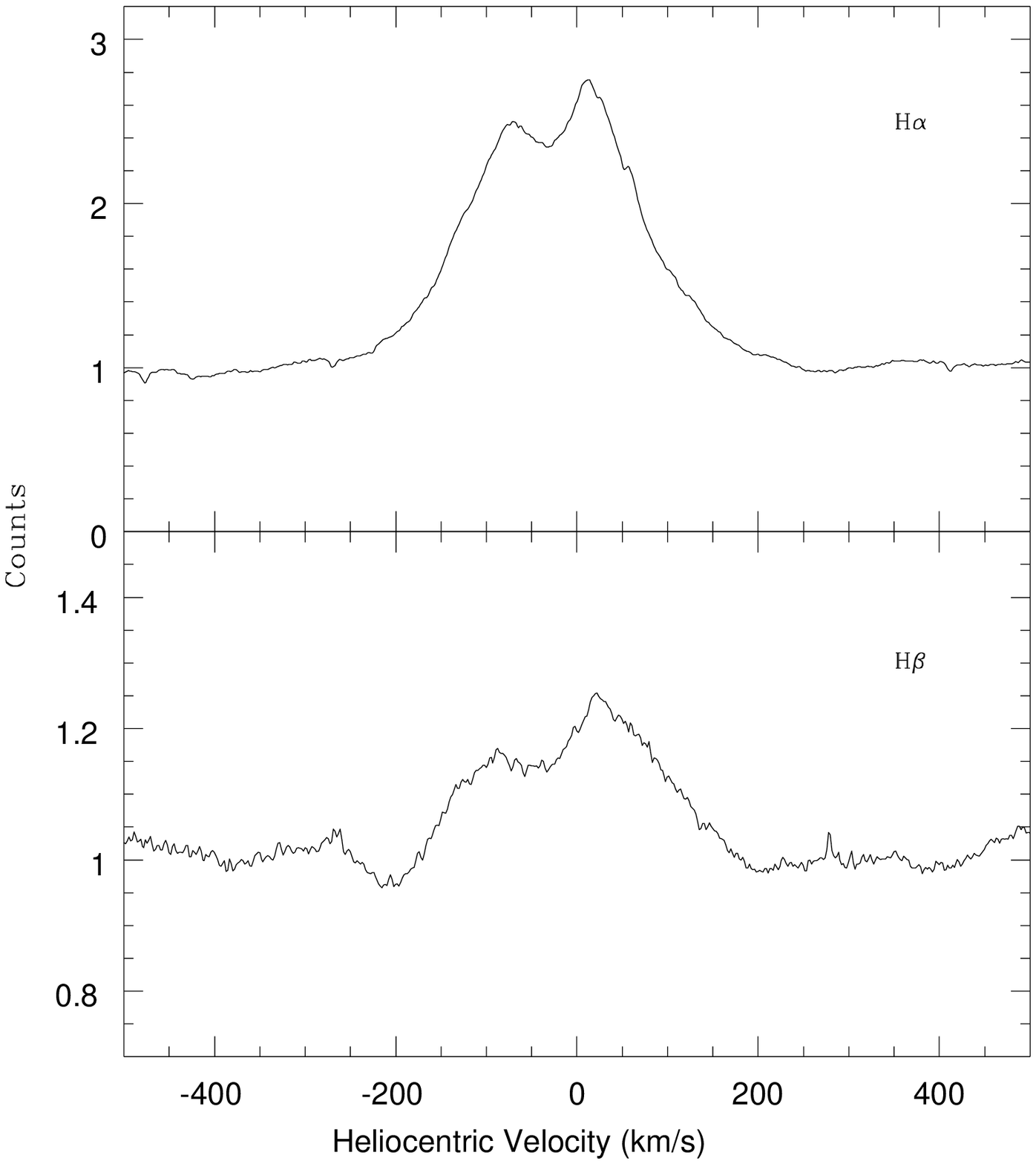]{H$\alpha$ and H$\beta$ profiles shown in the 
FEROS spectrum.
\label{fig4}} 

\figcaption[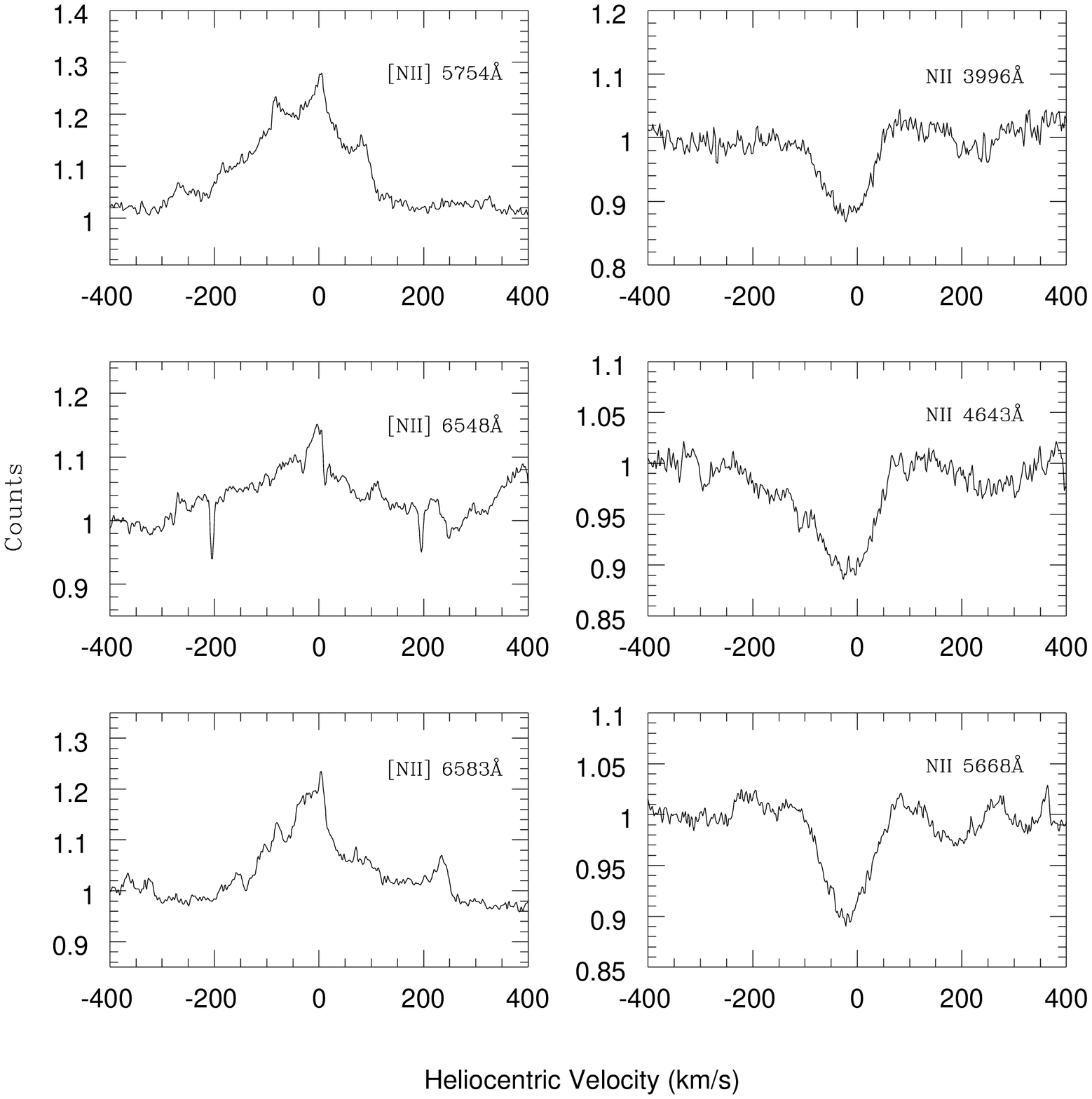]{Profiles of permitted and forbidden N\,{\sc ii} 
lines obtained from FEROS spectrum.
\label{fig5}} 

\figcaption[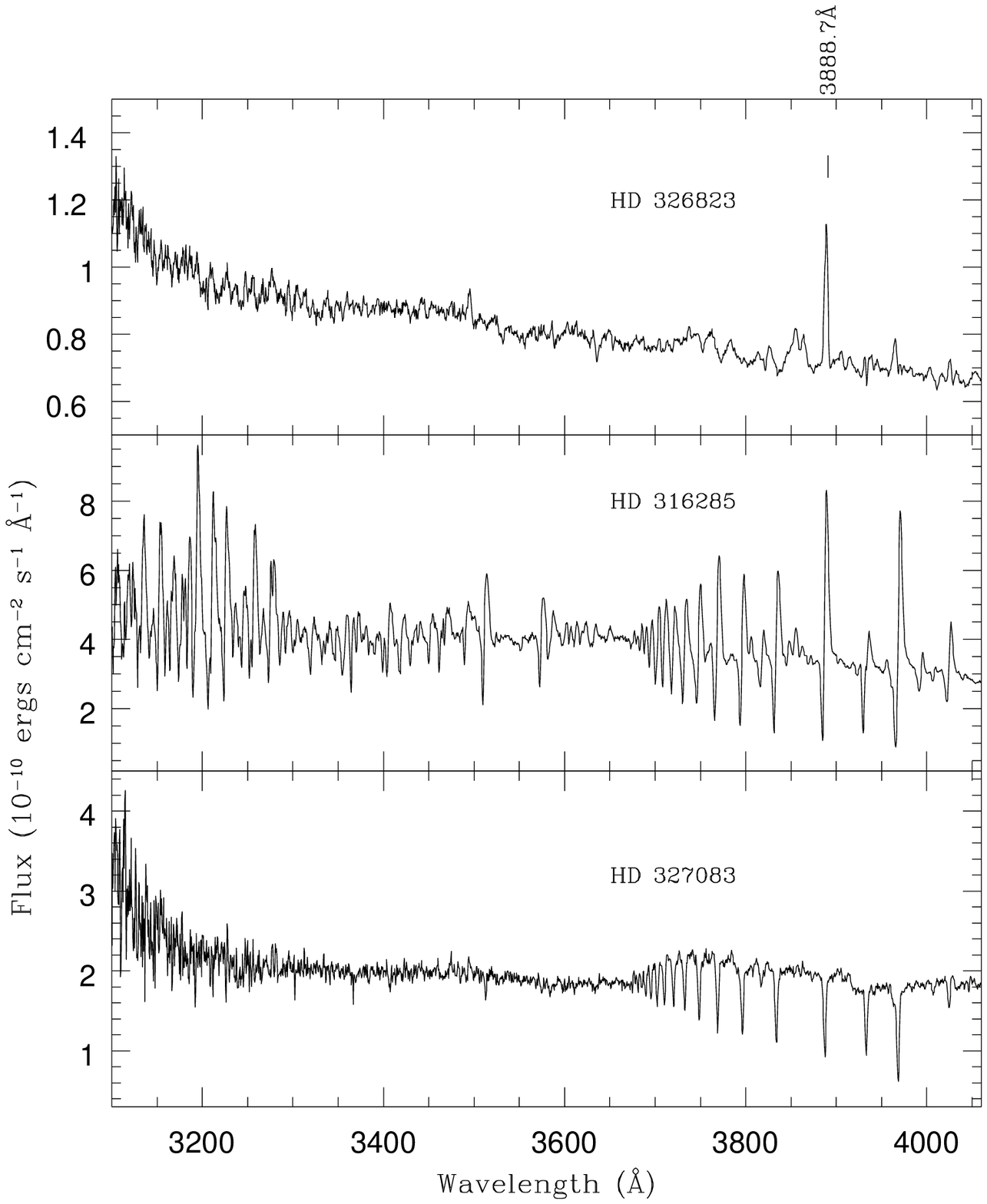]{Low resolution spectra of HD 326823, HD 316285, 
and 
HD 327083 with $\lambda$$_{central}$ $\sim$ 3600\AA.
\label{fig6}} 

\figcaption[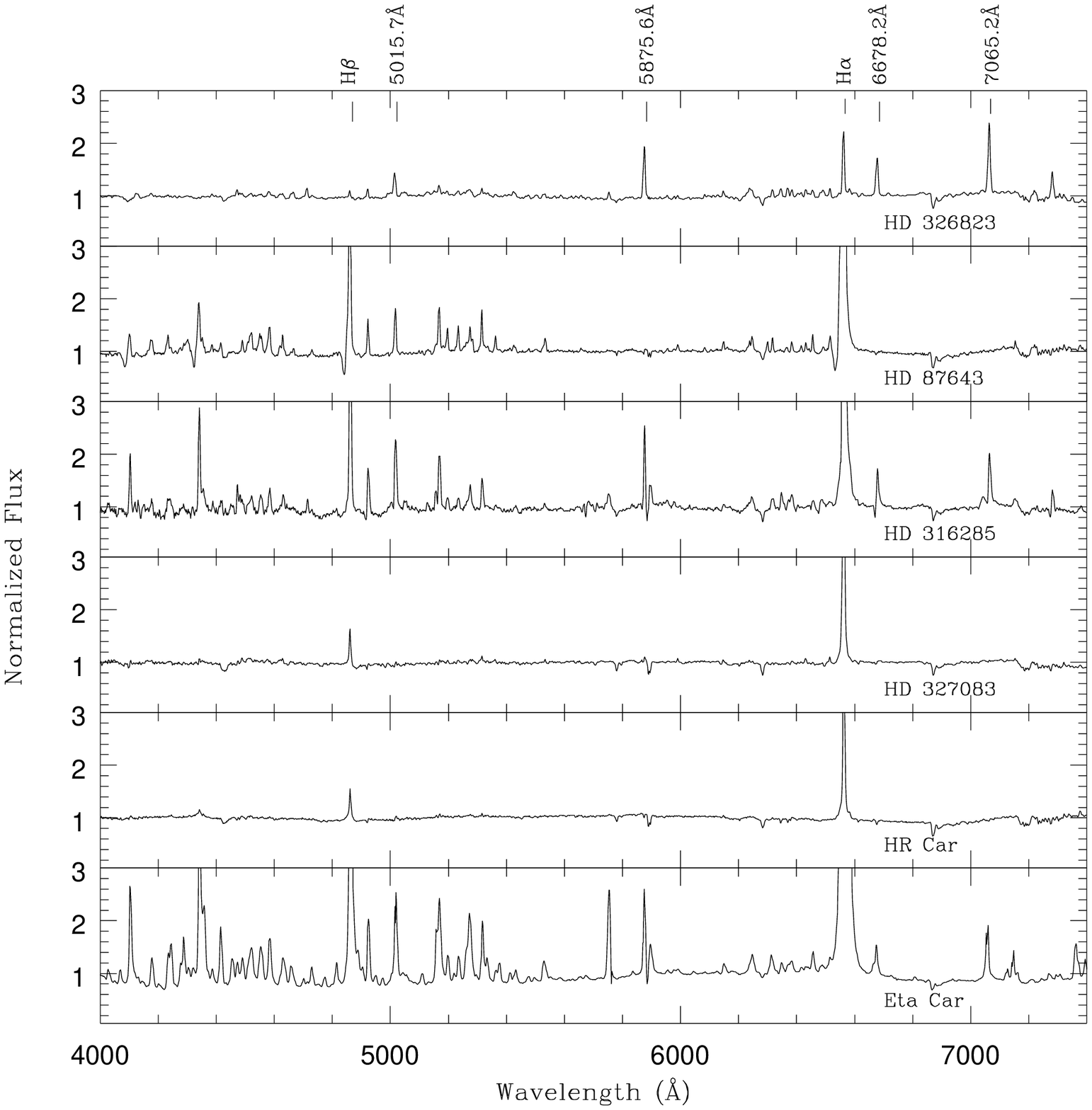]{Normalized low resolution spectra of HD 326823, 
HD 87643, HD 316285, HD 327083, HR Car, and $\eta$ Car.
\label{fig7}} 

\figcaption[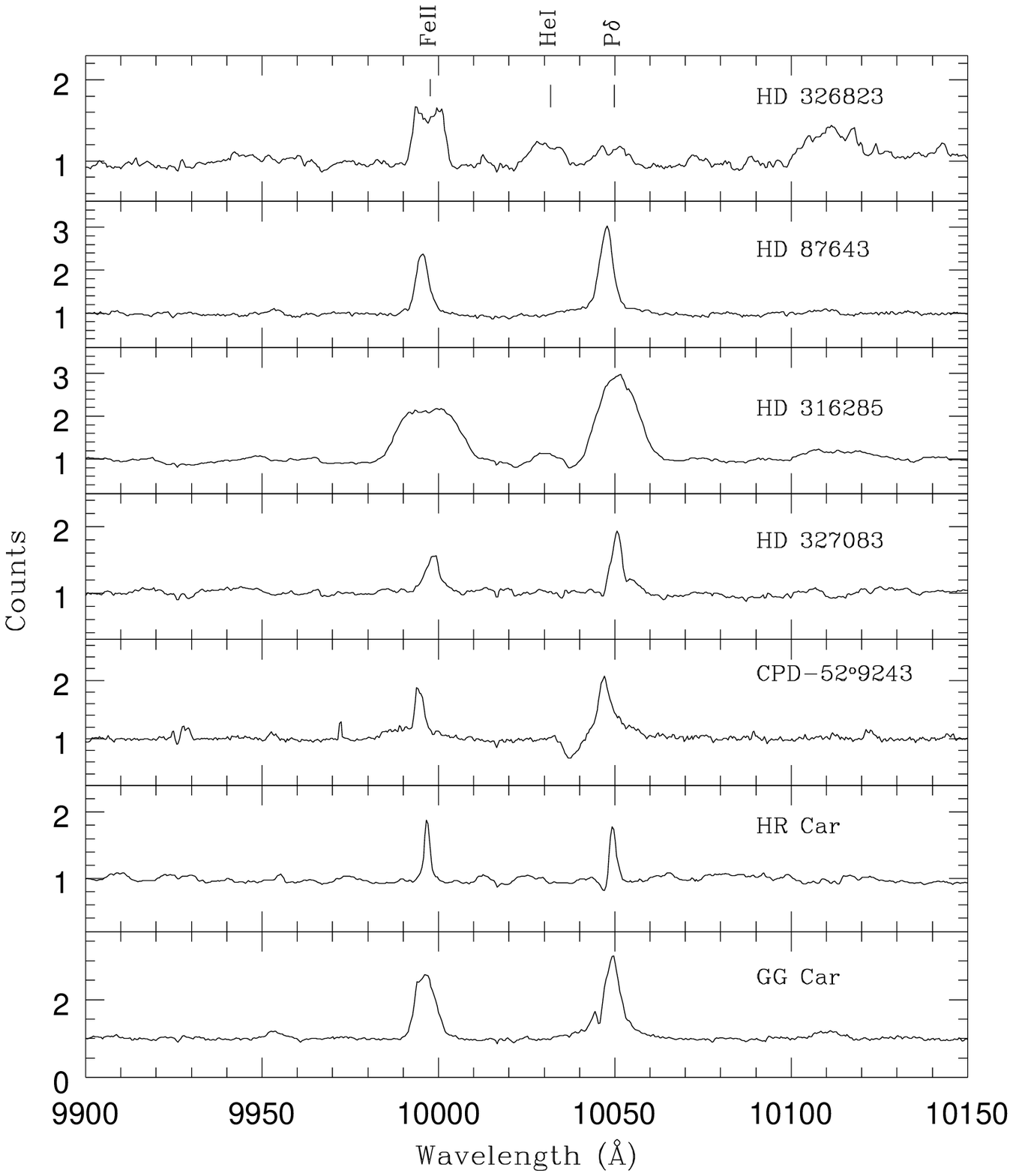]{Normalized Coud\'e spectra of HD 326823, HD 
87643, 
HD 316285, HD 327083, CPD-52$^o$9243, HR Car, and GG Car.
\label{fig8}}

\end{document}